\documentclass{article}
\usepackage{verbatim}
\usepackage{graphicx}
\usepackage{amssymb}
\usepackage{amsmath}
\usepackage{units}
\usepackage{setspace}

\newcommand{\cL}{{\cal{L}}}

\newcommand{\bld}[1]{\mbox{\boldmath $#1$}}

\newcommand{\bbe}{\bld e}

\newcommand{\bbs}{\bld s}

\newcommand{\bbx}{\bld x}
\newcommand{\bby}{\bld y}

\title{Modeling  spatial data using local likelihood estimation and a Mat\'ern to SAR  translation}

\author{
Ashton Wiens
\thanks{University of Colorado, Boulder, Colorado, USA}
\and  
Douglas Nychka
\thanks{Corresponding author: 
Department of Applied Mathematics and Statistics, 1500 Illinois St, Golden, CO 80401 
}
\and 
William Kleiber\footnotemark[1]
}

\begin{document}

\maketitle

\begin{abstract}

Modeling data with non-stationary covariance structure is important to represent heterogeneity in geophysical and other environmental spatial processes. In this work, we investigate a multistage approach to modeling non-stationary covariances that is efficient for large data sets. First, we use likelihood estimation in local, moving windows to infer spatially varying covariance parameters. These surfaces of covariance parameters can then be encoded into a global covariance model specifying the second-order structure for the complete spatial domain. The resulting global model allows for efficient simulation and prediction. We investigate the non-stationary spatial autoregressive (SAR) model related to Gaussian Markov random field (GMRF) methods, which is amenable to plug in local estimates and practical for large data sets. In addition we use a simulation study to establish the accuracy of local Mat\'ern parameter estimation as a reliable technique when replicate fields are available and small local windows are exploited to reduce computation. This multistage modeling approach is implemented on a non-stationary climate model output data set with the goal of emulating the variation in the numerical model ensemble using a Gaussian process.
\end{abstract}

\section{Introduction}
Climate models produce high-dimensional spatial fields of variables related to various processes that comprise the Earth system. 
To quantify uncertainty, ensembles are generated typically by perturbing initial conditions to the climate model; however the ensemble size is usually limited to a handful of members due to the extreme computational demands of such codes.
An alternative approach is to emulate the climate model output using a statistical model from which uncertainty can be readily derived. 
Many climate fields appear to be well approximated by a Gaussian process but the covariance structure is distinctly non-stationary.
This paper focuses on statistical emulation of high-dimensional climate model spatial output that exhibit substantial non-stationarity in  the spatial covariance. 
The major obstacles are in specifying a flexible non-stationary model that is amenable to estimation for large datasets, but also allows for computationally efficient simulation. 

We investigate a multistage approach to estimating and modeling non-stationary covariances, similar to the methodology in \cite{nychka2018modeling}. 
First, assuming the field is approximately locally stationary, we perform moving window local likelihood estimation to infer spatially varying Mat\'ern covariance parameters. 
In the approach of  \cite{nychka2018modeling}  these parameters  are mapped into those of the LatticeKrig 
model \cite{nychka2015multiresolution}. A more direct approach for regularly spaced observations is to exploit a relationship between the 
Mat\'ern parameters and those of a spatial autoregressive (SAR) random field, 
to  reproduce local Mat\'ern 
correlations. 
Finally, the spatially varying parameters are encoded into a global SAR precision matrix, specifying the global field's dependence structure simultaneously. 
As this work is at the intersection of non-stationary modeling, local estimation and Mat\'ern-SAR connections, we begin with a brief review on these distinct topics.

There are many general classes of non-stationary models, such as deformation methods \cite{sampsonguttorp, anderes2008estimating}, basis function methods \cite{cressie2008fixed, katzfuss2011spatio, nychka2015multiresolution, nychka2002multiresolution}, process-convolutions \cite{higdon1998process, higdon1999non, higdon2002space, paciorek2004non-stationary, fuentes2001new, fuentes2002spectral, zhu2010estimation}, and the SPDE approach \cite{lindgren2011explicit, lindgren2007explicit, simpson2012think, rue2005gaussian}. 
See \cite{risser2016non-stationary} for a review of non-stationary models and \cite{heaton2017methods} for a review of methods for large spatial data sets. 
Here, we investigate two existing non-stationary models from the GMRF families, and focus our attention on computationally efficient estimation using localized moving windows.

Local estimation is a well-established idea in spatial statistics \cite{haas1990kriging, haas1990lognormal, ver2004flexible, risser2015local}, and is popular in that full likelihood-based calculations, which become prohibitive for large sample sizes, can be circumvented. 
Moreover, this strategy is an easily parallelizable problem which can lead to further computational improvements. 
In practice, there is often no clear indication of which parameters in the model should vary spatially \cite{fuglstad2015does} and what spatial scales are appropriate for the parameter surfaces.                                                                                                                                                                                                                                                                                                                                                                                                                                                                      
This difficult modeling choice is avoided when using local estimation: we can allow all parameters to vary initially, and the local estimates will lead to diagnostics as to whether the parameters should be constrained to be constant or vary over space. 
Furthermore, with local estimation, we do not have to decompose the parameter functions into some prespecified low-dimensional representation as in \cite{fuglstad2015exploring, risser2016non-stationary, marques2019}, which can bias estimates. We should note that for single realizations of a spatial field and moderate sample sizes  a basis function expansion, such as the  recent work by
 \cite{marques2019}, is perhaps the most efficient  way to capture nonstationarity. 
Weighted local likelihoods have been studied to accommodate irregularly spaced observations \cite{anderes2011local}, but in this work we use a simple moving window applied to data on a lattice. 
Here, we focus on estimation of locally-varying Mat\'ern parameters, primarily because of their interpretability and to exploit a relationship between Mat\'ern and SAR covariance models, detailed below. 
To justify local estimation as a reliable technique, we use a Monte Carlo experiment to study the robustness of local estimation of the correlation range parameter. 

With locally estimated covariance parameters in hand, some care is required to combine these into a valid global non-stationary covariance model. 
A simple option is to use the estimates to construct local covariance functions and perform local simulation; however it is not clear how to stitch together independently simulated localized fields. 
In this work, we use a non-stationary spatial autoregressive (SAR) model, related to the Gaussian Markov random field (GMRF) approach to approximating GPs. 
The idea is to identify members of the Mat\'ern family of spatial processes as solutions to a stochastic partial differential equation. 
The SPDE is then discretized to a lattice and this motivates the form of the SAR \cite{lindgren2011explicit}. 
The correspondence between the Mat\'ern/SPDE form and a SAR was presented in \cite{lindgren2011explicit}, and an analytical formula was proposed to connect the parameters between the continuous and discrete cases. 
We have found that the analytical formula is inaccurate for large correlation ranges and one contribution of this work is to sharpen this relationship using numerical results. 
The advantage is that if we can successfully translate the Mat\'ern formulation into a SAR framework, we can exploit sparse matrix algorithms for fast computation.

An important contribution of this work is showing non-stationary spatial processes can be modeled by combining local maximum likelihood estimation with a simple global non-stationary covariance model that is straightforward to implement. 
As an illustration, we apply this multistage modeling framework to analyze a climate model output pattern scaling problem consisting of temperature anomaly fields from the Community Earth System Modeling's Large Ensemble project \cite{kay2015community}. 
These data cover about 13,000 spatial locations over the Americas, and exhibit strong nonstationarities that challenge the construction of statistical emulators.

\section{Stationary covariance models}
\label{s:2}

In this section, we discuss the connection between the isotropic Mat\'ern family of covariance models for Gaussian processes with the spatial autoregression (SAR)  construction for Gaussian Markov random fields (GMRFs), followed by an exploration of this relationship in a numerical study. 

\subsection{The Mat\'ern covariance model}
\label{ss:2.1}

Let $f(\bbx)$ be a mean zero Gaussian process on $\bbx\in{\mathbb R}^2$ with covariance function $k( \bbx, \bbx^\prime)\in{\mathbb R}$.
The Mat\'ern family of stationary covariance models is important because of its flexibility and the interpretability of its parameters. The  Mat\'ern covariance function with a unit range parameter is
\[
  C( \, d \, | \, \nu,  \sigma^2 \, ) =
  \sigma ^{2}{\frac {2^{1-\nu }}{\Gamma (\nu )}}{( d)}^{\nu }
  \mathcal{K}_{\nu }{( d  )},
\]
where $d$ is the Euclidean distance between  $\bbx$ and $\bbx^\prime$,  $\mathcal{K}_{\nu }(\cdot)$ is the modified Bessel function of the second kind of order $\nu$, and $\Gamma(\cdot)$ is the gamma function. $\sigma^2$ is the spatial process variance, and $\nu$ is the smoothness parameter which controls the mean square differentiability of the process. 
The isotropic covariance function with range  parameter $\kappa$ is given by 
\[
  k( \bbx, \bbx^\prime) = C(\kappa d\,|\,\nu,\sigma^2).
\]

\subsection{The SAR model}
\label{ss:SAR}

In contrast to modeling a covariance function for a process of continuous spatial variation, the SAR model parameterizes the precision matrix for the process on a discrete lattice. 
For the following development we denote by $\mathbf{y}$ a Gaussian process on an infinite regular lattice in ${\mathbb R}^2$. 

Denote by $y_{i,j}$ the element of $\bby$ at lattice location $(i,j)$.
A simple isotropic SAR model can be written using lattice notation as 
\begin{equation} \label{e:2}
  \begin{array}{c|c|c}
    0 & -1 & 0 \\
    \hline
    -1 & 4+ \kappa_S^2 & -1 \\
    \hline
    0 & -1 & 0 \\
  \end{array}
\end{equation} 
which visually illustrates a set of decorrelating weights on the random vector $\bby$. 
To be precise, we interpret (\ref{e:2}) as implying that the following equation holds 
\begin{equation} \label{e:2.1}
  (4+\kappa_{S}^2) y_{ij} - ( y_{i-1,j} + y_{i+1,j} + 
  y_{i,j-1} + y_{i,j+1} ) = e_{ij},
\end{equation} 
for a mean zero unit variance normally distributed white noise vector $\bbe$ with components registered to the spatial grid $\{e_{ij}\}_{i,j}$.

Here $\kappa_{S}> 0 $ is suggestive of a range parameter controlling the correlation length scale dependence of the field and is similar, but not identical, to $\kappa$ for 
the continuous Mat\'ern case above. 
For the model in (\ref{e:2}), one can populate a matrix $B$ using (\ref{e:2.1}) such that $B\bby =  \bbe$. 
Furthermore, it is clear that if $\kappa_{S}> 0$, the diagonal dominance of $B$ guarantees its invertibility. 
With $\bby  =   B^{-1} \bbe$, the covariance matrix for $\bby$ is $ B^{-1}B^{-T}$ and thus $\bby$ has precision matrix $Q= B^T B $. Moreover, it is straightforward to see that the precision matrix implied by the SAR model is sparse. 
The sparsity property makes the SAR model amenable to modeling large data sets because the sparse precision matrix can be used in place of a dense covariance matrix for likelihood estimation and simulation. 
Following the ideas from \cite{lindgren2011explicit} one can iterate the spatial autoregressive weights to obtain higher order models that approximate smoother processes. 
For example, if $ B B \bby = \bbe $, this implies a SAR model extending to second-order nearest neighbors and gives the precision matrix: $Q_2= (B B)^T (B B)$.   
The SAR model detailed here is a special case of a GMRF. 
For a given row of the precision matrix, the nonzero, off-diagonal entries index the neighbors that determine the Markov property. 
Citing the \textit{order} of neighborhoods can cause some confusion depending on whether one is referring to  SAR weights , or precision, matrix.
For the first-order SAR described above, the nonzero elements in $Q$ will include second-order neighbors. 
A first-order SAR will be a GMRF based on second-order neighbors and the weights will depend on $B$.
Two additional points concerning the SAR model are important.
First, for a given dataset $\bby$ is not defined on an infinite lattice, but rather a (typically) rectangular one; this stencil of (\ref{e:2}) should be modified at the boundaries of the domain. 
The center value of the stencil should be the $\kappa_S^2$ minus the sum of the weights of its non-zero neighbors, although a simpler option is to artificially extend the lattice a few nodes outwards to reduce boundary effects. 
Second, the value of $\kappa_S$  is one-to-one functionm for  the marginal variance of the process.

A version of the SAR model that exhibits approximate stationarity and also geometric anisotropy will be detailed in Section \ref{s:3}. 

\subsection{The Mat\'ern-SAR link}

Lindgren and Rue \cite{lindgren2011explicit} developed an approximation of Gaussian random fields with Mat\'ern covariance functions using GMRFs with particular SAR structures. 
The connection is established through a stochastic partial differential equation (SPDE) formulation in that a Gaussian field $u(\mathbf s)$ with stationary Mat\'ern covariance is a solution to the SPDE 
\[
  (\kappa^2 - \Delta)^{\nicefrac{\alpha}{2}} u(\mathbf s) = \mathcal{W}(\mathbf s)
\]
where $\alpha = \nu + \frac{d}{2}$, $\kappa > 0, \nu > 0$, $\mathbf s \in \mathbb{R}^d$, $d = 1$ or $2$, and $\mathcal{W}(s)$ is a generalized white noise process with zero mean and variance $\sigma^2$. 
As in the Mat\'ern model, $\nu$ controls the smoothness of realizations of the Gaussian field. 
Fixing $\nu = 1$ and $d = 2$, \cite{lindgren2011explicit} showed that the SAR covariance structure obtained by discretizing the pseudodifferential operator $(\kappa^2 - \Delta)$ approximates a Mat\'ern covariance structure with range $\kappa \approx \kappa_S$; this relationship is approximate. 
Similar results can be obtained for different smoothness parameters $\nu$ by convolving the finite difference stencil in (\ref{e:2}) with itself $\nu$ times, as detailed in the previous section for $\nu=1$ and $\nu=2$.

\subsection{Numerical translation of range parameters between the Mat\'ern and SAR models}
\label{ss:1}

The connection between the isotropic Matern family and a SAR relies on the approximation of a discretized Laplacian operator with finite differences of the fields on a lattice.  
To develop an accurate statistical model, it is important to quantify the error in such an approximation and improve its calibration over the limiting expression suggested in \cite{lindgren2011explicit}. 
In this section we provide numerical evidence to show that an accurate calibration is possible if restricted to specific ranges of the covariance parameters. 

Our calibration setup is as follows: given a Mat\'ern range parameter $\kappa$, we estimate the value of $\kappa_S$ in the SAR model which gives the best approximation to the Mat\'ern correlation function.  
We consider the smoothness of the Mat\'ern model fixed at $\nu=1$ and $\nu=2$, and with unit marginal variance for all models. 
The first step is to fix the Mat\'ern range parameter and evaluate a Mat\'ern correlation matrix for the process evaluated on the lattice grid. 
Then, we find the optimal $\kappa_S$ from the SAR model by computing its equivalent correlation matrix (the standardized inverse precision matrix) and minimizing distance between the Mat\'ern and SAR correlation matrices.

It is known that the SAR covariance model suffers from edge effects. 
To avoid the interference of edge effects, we quantify the difference between the two correlation matrices by only comparing the correlations from the central lattice point under both models. 
For an $N \times N$ lattice of locations, with $N$ odd, let $\boldsymbol \sigma_{\kappa}$ denote the vector of correlations between the center point in this lattice and all other locations based on the Mat\'ern covariance function with range $\kappa$.  
Let $\boldsymbol \sigma_{\kappa S}$ be the analogous correlation vector for the SAR model with range parameter $\kappa_S$.  
We then find 
\[
  \hat{\kappa}_S( \kappa ) = \underset{\kappa_S}{\operatorname{argmin}} \| \boldsymbol \sigma_\kappa - 
  \boldsymbol \sigma_{\kappa S} \|_2.
\]
If the relationship proposed by \cite{lindgren2011explicit} holds we would expect $\hat{\kappa}_S(\kappa) \approx \kappa$.

\begin{figure}
    \centering
    \makebox[0pt]{
        \includegraphics[width=5in]{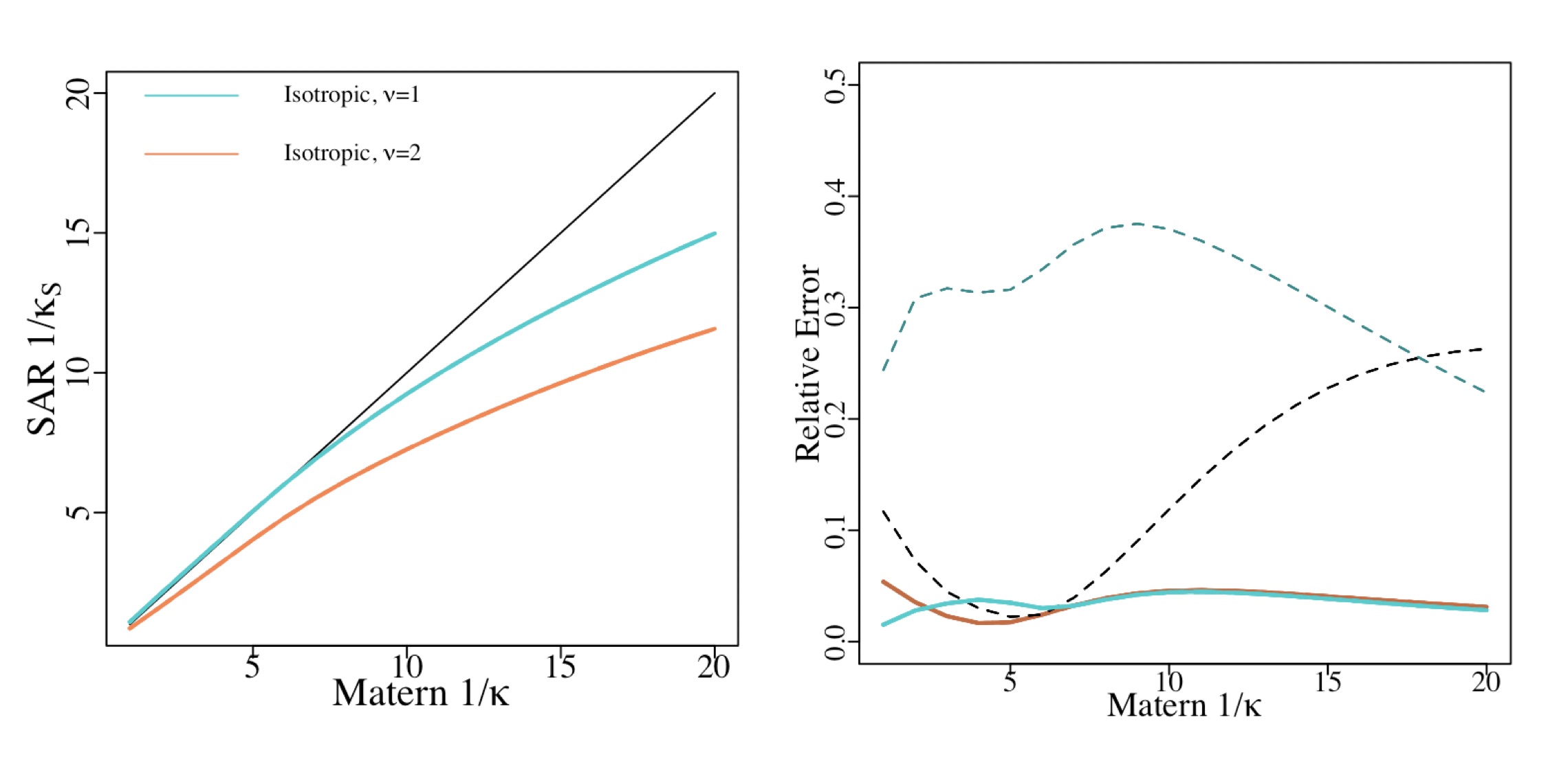}  
    }
    \caption{For the isotropic case, the optimal $1/\kappa_S$ parameter for a given Mat\'ern inverse range $1/\kappa$ is plotted in (a). The relative error incurred by using the SAR model with optimal $\kappa_S$ as an approximation to the Mat\'ern model is shown in (b) (estimated relationship in solid lines, \cite{lindgren2011explicit} relationship of $\kappa_S=\kappa$ in dashed lines).}
    \label{f:1}
\end{figure}

$\kappa^{-1}$ is varied over the interval $[1, 20]$ with $N=51$ and  lattice points having unit spacing.  
The approximation results are summarized in Figure \ref{f:1}. 
In \ref{f:1}(a) $\hat{\kappa}_S( \kappa )^{-1}$ is plotted as a function of $\kappa^{-1}$. 
Orange corresponds to the $\nu=1$ case, cyan corresponds to $\nu=2$, and the solid black line shows the theoretical relationship, $\kappa_S^{-1} = \kappa^{-1}$ from \cite{lindgren2011explicit}.  
From this experiment we conclude that, at least at this level of discretization, it is important not to rely on the analytic formula to translate between $\kappa$ and $\kappa_S$ parameters. 

The relative error of using the SAR correlation with $\kappa_S$ value derived from the numerical experiment is shown in \ref{f:1}(b). 
The $\ell_2$ distance measure used in the optimization of the model correlation matrices is used to quantify the resulting model error, normalized by $\| \boldsymbol \sigma_{\kappa S} \|$. 
The relative error incurred when using $\frac{1}{\kappa_S} = \frac{1}{\kappa}$ as shown by the dashed lines can be on the order of $20\%$ or worse for higher correlation ranges. However, using the calibrated range (solid lines) gives relative errors on the order of $5\%$.

\section{Anisotropic and non-stationary covariance models}
\label{s:3}

In this section, we extend the Mat\'ern and SAR models to include geometric anisotropy, conduct the numerical study for this case, and finally discuss some related non-stationary models.

\subsection{The anisotropic Mat\'ern covariance model}

The Mat\'ern family can be extended to include geometric anisotropy by defining a distance measure based on a linear scaling and rotation of the coordinates.
Let 
 $A  =  D^{-1} U^T$ be a  $2\times2$ matrix 
where $U$ is a rotation matrix parameterized by angle $\theta$
\[
  U = \begin{bmatrix} \cos(\theta) & -\sin(\theta) \\ 
  \sin(\theta) & \cos(\theta) \end{bmatrix},
\]
and
\[
  D = \begin{bmatrix} \xi_{1} & 0 \\ 0 & \xi_{2} \end{bmatrix}
\]
is a diagonal matrix scaling the $s_1$ and $s_2$ coordinate axes separately. 
Then the pairwise Mahalanobis distance between two locations $\bbs,\bbs'$ is defined as
$ d= \| A\bbs - A\bbs' \| $
which is used as the argument to the isotropic Mat\'ern covariance function. 
A useful interpretation of this form is that if one transforms the coordinates according the linear transformation $A$ then the resulting field will be isotropic. 

\subsection{The anisotropic Mat\'ern-SAR link}

The SAR model can also be extended to incorporate geometric anisotropy. Let $H$ denote a $2\times 2$ symmetric positive definite anisotropy matrix and modify the Laplacian in the pseudodifferential operator as follows
\begin{equation} \label{SPDE}
  (\kappa^2 - \nabla \cdot H \nabla)^{\nicefrac{\alpha}{2}} u(\mathbf s) = 
  \mathcal{W}(\mathbf s).
\end{equation} 
To avoid potential ambiguity it is helpful to  identify the Laplacian operator above for two dimensions in its expanded form as 
\[
  \nabla \cdot H \nabla  \equiv  H_{1,1} \frac{\partial^2}{\partial^2 s_1} +  
  2 H_{2,1} \frac{\partial^2}{\partial s_1 \partial s_2} + 
  H_{2,2} \frac{\partial^2}{\partial^2 s_2}.
\]
From this expression, a first-order finite difference discretization of the anisotropic SPDE 
at  (\ref{SPDE}) gives the following stencil for filling the rows of the $B$ matrix, 
\begin{equation} \label{e:3}
  \arraycolsep=6.0pt\def\arraystretch{2.5}
  \begin{array}{c|c|c}
    \frac{2H_{12}}{h_{s_1} h_{s_2}} & -\frac{H_{22}}{h_{s_2}^2} & -\frac{2H_{12}}{h_{s_1} h_{s_2}} \\
    \hline
    -\frac{H_{11}}{h_{s_1}^2} & \;\; \kappa^2 + \frac{2H_{11}}{h_{s_1}^2} + 
    \frac{2H_{22}}{h_{s_2}^2} \;\; & -\frac{H_{11}}{h_{s_1}^2} \\
    \hline
    -\frac{2H_{12}}{h_{s_1} h_{s_2}} & -\frac{H_{22}}{h_{s_2}^2} & \frac{2H_{12}}{h_{s_1} h_{s_2}} \\
  \end{array} 
\end{equation}
where $h_{s_1}$ and $h_{s_2}$ are the grid spacings along the x-axis and y-axis. This is just a reparameterization of the results in Appendix A of \cite{lindgren2011explicit} but facilitates the practical translation between these models. Moreover, setting $h_{s_1} = h_{s_2} =1$, $H_{12} = H_{21} =0$, and $H_{11} = H_{22} = 1$ yields the first-order isotropic model from (\ref{e:2}).

Finally, we connect the role of $H$ in the SPDE formulation to the anisotropic model for the Mat\'ern.  
Under the linear transformation $A=D^{-1}U^T$ from Section \ref{ss:2.1},  let $\bbs^* =  A^{-1}  \bbs $, and let $u$ be an isotropic field solution to the SPDE with Laplacian $\nabla \cdot \nabla$. Furthermore, set $u^*(\bbs) = u(A^{-1} \bbs) = u(\bbs^*)$. 
Then from elementary properties of the gradient 
\[
  \nabla u^*( \bbs) = \nabla \left( u( A^{-1} \bbs) \right) =  
  \left. A^{-1} \nabla u(A^{-1} \bbs) \right| _{\bbs= A \bbs^*} =  
  \left. A^{-1} \nabla u( \bbs^*) \right| _{\bbs^*= A^{-1} \bbs}  
\] 
and so we have  
\[ 
  \nabla \cdot \nabla u^*(\bbs) =  
  (A ^{-1}\nabla) \cdot A^{-1} \nabla u(\bbs^*)  =  
  \left. \nabla \cdot A^{-T} A^{-1} \nabla u(\bbs^*) \right| _{\bbs^*= A^{-1} \bbs}.
\] 
From this expression we identify $H= A^{-T}A^{-1}$.  
From Section \ref{ss:2.1}, if $u$ is an isotropic field then $u^*$ will be anisotropic with coordinates transformed by $A^{-1}$. 
Moreover, $u^*$ will also be the solution to the SPDE with $H= A^{-T}A^{-1}$. 
This connection provides guidance how to interpret $H$. 
Finally, note that if $A$ is a pure rotation then $H=I$ and isotropy is preserved.  

\subsection{Numerical translation of anisotropic range parameters between the Mat\'ern and SAR models}

\begin{figure}
    \centering
    \makebox[0pt]{
        \includegraphics[width=5in]{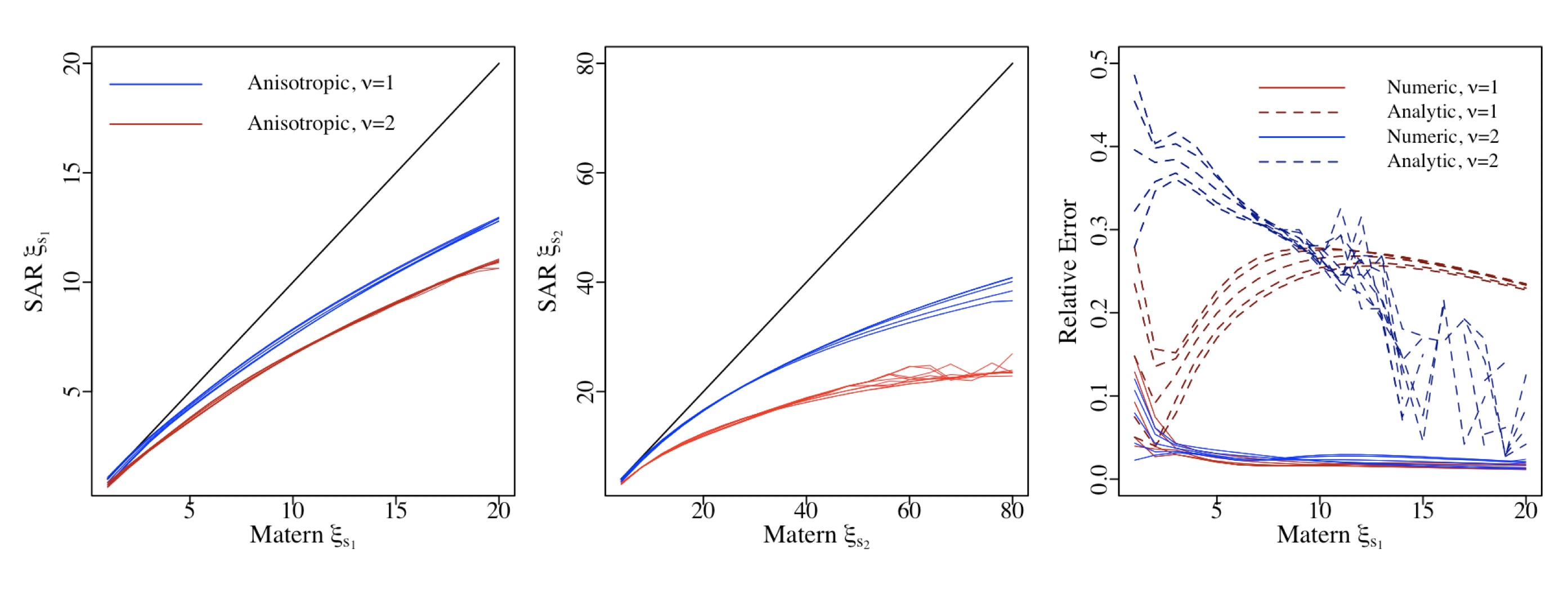}
    }
    \caption{For the anisotropic case, the optimal eigenvalues of $H$ are plotted against the fixed diagonal values of $D^2$ in panels (a) and (b), and the relative error is shown in (c), analogous to Figure \ref{f:1}.}
  \label{f:2}
\end{figure}

In the climate data analysis below, we find it necessary to include geometric anisotropy in the covariance model. 
For this reason, we also investigate how the presence of geometric anisotropy affects the numerical correspondence established in \ref{ss:1}. 
In this experiment, we encode fixed values for $\xi_{s_1}$ and $\xi_{s_2}$ in a Mat\'ern correlation matrix such that the length scale ratio $\xi_{1} \colon \hspace{-0.4mm}  \xi_{2} = 4 \colon \hspace{-1mm} 1$, which is consistent with the anisotropic estimates in the data analysis. 
In particular, we let $\xi_{1} = 1, \cdots, 20$ and $\xi_{2} = 4\xi_{s_1}$. 
Then, the optimal eigenvalues of the SAR anisotropy matrix $H$ are found. 
The experiment was repeated with rotation angles $\theta = 0^{\circ}, 10^{\circ}, \cdots, 90^{\circ}$, where the rotation angle is assumed to be known and fixed in the eigendecomposition of the $H$ matrix. 
The anisotropic parameter translation results for $\xi_{s_1}$ and $\xi_{s_2}$ are shown in panels (a) and (b) of Fig \ref{f:2}, respectively, and the relative error of approximation is shown in panel (c). 
Overall, the behavior of the approximation is similar to the isotropic case: the estimated eigenvalues of the SAR anisotropy matrix $H$ are smaller than the eigenvalues fixed in the Mat\'ern anisotropy $D^2$ matrix. 

The approximation results may be slightly affected by the rotation angle and oblateness of the geometric anisotropy, but the effect is negligible in practice. From these results, we have ascertained a numerical translation among the anisotropy parameters. We can use these results to translate locally estimated anisotropic Mat\'ern model parameters into SAR parameters with improved accuracy over the conjectured analytic relationship.

\subsection{A non-stationary SAR model}

A non-stationary SAR model can be constructed by allowing the parameters $\kappa, H$, and $\sigma^2$ in the generating SPDE to vary over space. 
Let
\[
  \cL(\bbs) = H_{1,1}(\bbs) \frac{\partial^2}{\partial s_1^2} +  
  2 H_{2,1}(\bbs)\frac{\partial^2}{\partial s_1 \partial s_2} + 
  H_{2,2}(\bbs) \frac{\partial^2}{\partial s_2^2}.
\]
The SPDE can then be written as
\[
  (\kappa^2(\mathbf s) - \cL(\bbs) )^{\nicefrac{\alpha}{2}} u(\mathbf s) = 
  \mathcal{W}(\mathbf s)
\]
where $\kappa(\mathbf s) > 0$, $\mathcal{W}(s) \sim \operatorname{WN}(0, \sigma^2(\mathbf s))$, and $\sigma^2(\mathbf s) > 0$.  
Furthermore, we specialize to a spatially varying linear transformation of the coordinates, $A(\bbs)$, and so  $H(\bbs) = A^{-T} (\bbs) A^{-1}(\bbs)$. 
Note that $A(\bbs)$ varying in space is equivalent to specifying spatial fields for $\theta$, $\xi_{s_1}$ and $\xi_{s_2}$ within $U$ and $D$. 

Discretizing this equation results in a valid GMRF that is non-stationary. In particular, the autoregressive $B$ matrix from Section \ref{ss:SAR} could have  different elements in each row based on the variation in $H(\bbs)$ or $\kappa(\bbs)$. However, $B$ will still be a sparse matrix and $Q= B^TB$ will always be positive-definite. 
The process variance can also be allowed to vary in the same way as with the non-stationary Mat\'ern model, but this must be done balancing the identifiability of $\kappa$ and $H$ and the fact that edge effects may introduce spurious variation in the  GMRF variance. Our approach is to first construct the precision matrix and then use sparse matrix methods to solve for the diagonal elements of the covariance matrix. The rows of $B$ are then weighted so that  this new version gives a GMRF with constant marginal variance. With this normalization of the SAR model $\sigma(\bbs)^2$ can be introduced to capture explicit spatial variation in the process marginal variance.

\section{Local moving window likelihood estimation}

\subsection{Local estimation strategy}

Estimating a non-stationarity model can be challenging due to the increased number of covariance parameters. When enough data is available, however, local estimation can give insight into what type of non-stationarity is present. Moreover, we illustrate in this section that a modest number of replicated fields results in stable local covariance estimates. 

Local estimation is usually accompanied by the assumption of approximate local stationarity. 
For this work, we define local stationarity and the local likelihood estimation technique for a Gaussian process with stationary Mat\'ern covariance as follows. 
First, divide the region of interest $\mathcal{D}$ into $M$ possibly overlapping subregions, or {\it windows},  $\mathcal{D}_1, \mathcal{D}_2, \cdots, \mathcal{D}_M$. 
Then the assumption of approximate local stationarity is that we can model the data $\bby_i$ within the subregion $\mathcal{D}_i$ using a stationary Gaussian process $Y_i$ defined using the following specification:
\begin{equation}  \label{e:4}
  Y_i(\mathbf s) = \mu_i(\mathbf s) + Z_i (\mathbf s) + \varepsilon_i (\mathbf s)    
\end{equation}
where $\varepsilon_i$ is mean zero spatial white noise with variance $\tau_i^2$, and $Z_i$ is a mean zero Gaussian process with anisotropic but stationary Mat\'ern covariance function, and $\mu_i(\mathbf s)$ is a fixed mean function.
Let $G_i = \text{Var}\,\bby_i$ be the spatial covariance matrix of $\bby_i$ for the $i$th region. 
Then the local log likelihood based on $p$ independent replicates $\{\bby_{ij}\}_{j=1}^p$ for the $i$th region is, up to a constant,  
\begin{equation}  \label{e:5}
  \frac{p}{2} \log | G_i |^{-1} - 
  \sum_{j=1}^p \frac{1}{2}(\bby_{ij} - \boldsymbol \mu_i )^T G_i^{-1} (\bby_{ij} - \boldsymbol \mu_i)
\end{equation}

where $\boldsymbol{\mu}_i$ is the mean function $\mu_i(\mathbf s)$ evaluated at the locations of $\bby_i$. 
The likelihood is approximate because we are assuming stationarity within each data window. 

After partitioning the data, finding each local likelihood estimate is an embarrassingly parallel task,
 which makes it a viable strategy for large data sets using many computational cores. In fact, in our application the parallelization is efficient to the point that we take  the subregions to be an exhaustive set of moving windows centered at every grid point. 
We assign these estimated parameters to the location of the center of the subregion $\mathcal{D}_i$, and
after translating into the SAR parameterization these parameters specify the row of $B$, the SAR matrix, at this location. This assignment is, of course, predicated on the assumption that over the region there is little variation in these parameters. This issue will be discussed in more detail in the last section. 

Given that the SAR model also gives a specification of the covariance it may seem indirect that the local estimates focus on the Mat\'ern model, and then the estimates are transformed into the SAR representation. 
An alternative would be to estimate the SAR version directly from local likelihood windowing. There are several reasons for the two-step procedure. 
Fitting the covariance model directly avoids any boundary effects that would come about by applying the SAR to a small window. 
Furthermore, the Mat\'ern parameters are easier to interpret and will be more simple to model in a hierarchical statistical framework. 

\subsection{A numerical study of local Mat\'ern estimation}
\label{ss:2}

A practical issue for a local approach, especially in the context of determining covariance parameters, is whether the number of replicates and the size of the window are adequate for robust estimation of parameters. 
Although choosing a data adaptive window is beyond the scope of this work, it is important to identify the conditions under which parameter estimates will be accurate. 
Moreover, it is also useful to understand the benefits of replicate spatial fields in estimating a covariance model. 
In particular, the hope is that replication makes it possible to estimate correlation ranges that are much larger than the local window size. 

\begin{figure}[h]
    \centering
   \includegraphics[scale=0.12]{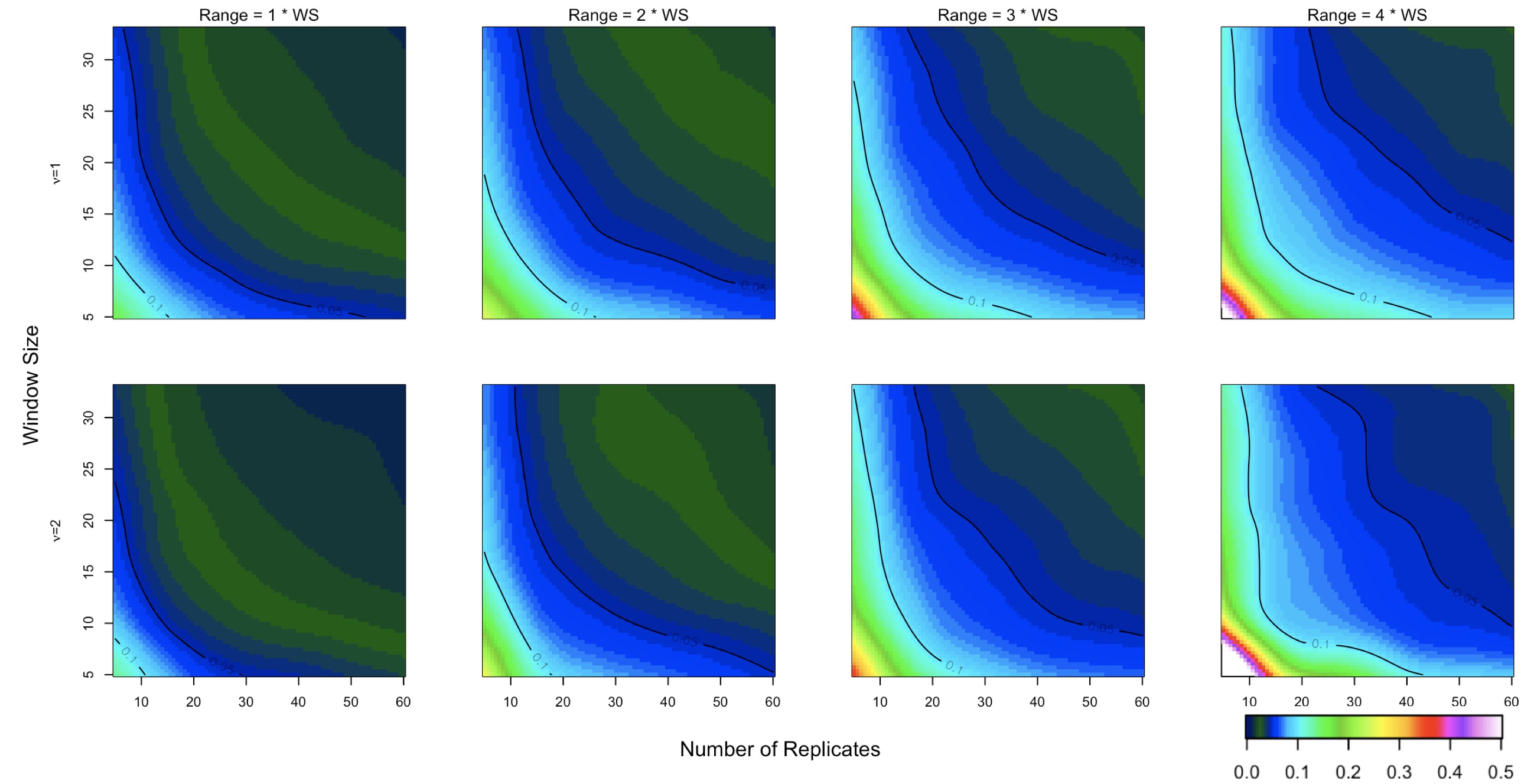}
    \caption{Each panel displays the absolute percent error from estimating the Mat\'ern range parameter given a certain number of replicates and a window size (size of grid). Fixed Mat\'ern range parameters one, two, three, and four times the size of the grid were tested, faceted in columns (a)-(d). The top row corresponds to $\nu=1$ and bottom to $\nu=2$. Thin plate splines were fit using the 100 repeated optimization results, performed at each grid location. The splines were used to predict the surfaces shown. Note that white indicates $>50\%$ error}
    \label{f:3}
\end{figure}

We perform a Monte Carlo experiment with four factors: window size ranging between a $5 \times 5$ grid and a $33 \times 33$ grid, the Mat\'ern range parameter being multiples of $1,2,3,$ and $4$ times the window size, the Mat\'ern smoothness parameter taking on values $1$ and $2$, and the number of replicates ranging between 5 and 60.  
Thus the full factorial design is $11 \times 4 \times 2 \times 9$ (window size $\times$ range parameter $\times$ smoothness parameter $\times$ replicates). 
For each combination, replicates with given range and smoothness were simulated with given window size, and the Mat\'ern range was estimated using maximum likelihood. 
This was done 100 times for each combination, and statistics were assembled from the 100 independent maximum likelihood estimates (MLEs) for the range parameter. 
The main quantity of interest is the percent error of the estimate, and so percent error surfaces as a function of replicate number and window size are summarized in Figure \ref{f:3}.  
In this experiment, the range parameter was varied based on the window size which may seem unusual.  
However, the motivation was to address the computational requirements of the problem: given a computational budget to accommodate windows of a specific size, what size range parameter can be accurately estimated? 
Note that with constraints on the window size, accuracy can also be improved by increasing the number of replicates. 

The surfaces in Figure \ref{f:3} can be used as guidelines to decide how many replicates are necessary and what window size should be used to achieve a specific estimation error tolerance, given something is known about the size of the range to be estimated. 
The results are encouraging: only a small number of replicates ($>10$) are needed with a window size of $>10$ to estimate a range of 10. 
In the extreme case, a Mat\'ern range four times the size of the window might be estimated to within $10\%$ error if 30 replicates are available and using a window size of 10 or greater. 
Using these guidelines, we can be more confident that local moving window likelihood estimation is a viable technique if enough data is used.

\section{Climate data application}

The data set from the CESM Large Ensemble project \cite{kay2015community} is comprised of 30 spatial fields that can be assumed to be independent replicates from the same distribution. 
This feature is based on the nature of climate model experiments run over a long period and started with different initial conditions. 
\cite{nychka2018modeling} first analyzed these data using the LatticeKrig model, and the original article details the climate science application. 
The data locations are on a $288 \times 192$ grid with approximately one degree resolution, covering the entire globe. The specific data set used in this application is publically available  in R binary format from  the LatticeKrig github repository \footnote{ See {\tt github.com/NCAR/LatticeKrig/Datasets/LENNS/BRACEUfields.rda} but also refer to the README file in this folder for more background in using these data. }
Details about the pattern scaling approach to statistical emulation can also be found in \cite{alexeeff2018emulating}. 
Briefly, each field is a measure of how the local temperature average is affected by a global temperature average increase of one degree Celsius. 
Generating this ensemble requires supercomputer resources. 
The statistical task is to represent these spatial fields with a probability distribution where it is more efficient to generate additional fields (e.g. several hundred or thousands) that track the original 30 member model results. 
\begin{figure}[t]
  \centering
  \makebox[0pt]{
    \includegraphics[width=5in]{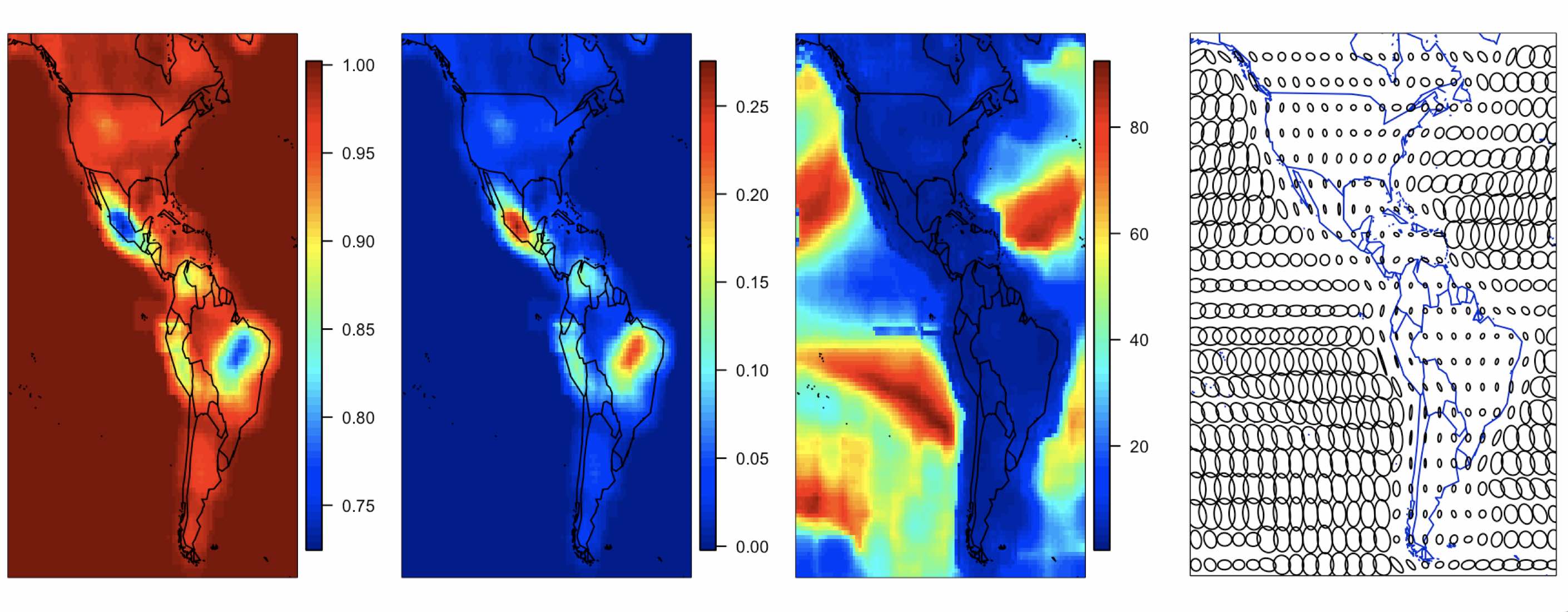}
  }
  \caption{Fitted parameter fields based on the moving window likelihood estimation. 
  The (a) variance $\sigma^2( \mathbf s)$, (b) nugget $\tau^2( \mathbf s)$, 
  (c) geometric average range $\sqrt{\xi_{1} ( \mathbf s) \xi_{2} ( \mathbf s)}$, 
  and (d) anisotropy ellipses defined by $A^T ( \mathbf s)A ( \mathbf s)$.}
  \label{f:moving}
\end{figure}

We focus on the subregion including the Americas and surrounding oceans containing $13,052$ spatial locations on a $102 \times 128$ grid. 
The top row of Figure \ref{f:4} shows the first four sample fields from the data set we analyze. The one data modification from \cite{nychka2018modeling} is that, in addition to subtracting the ensemble mean from each grid box, we have also standardized the fields by dividing by the ensemble standard deviation of each grid box.

\subsection{Covariance parameter estimates}

We experiment with moving window local MLEs using window sizes between $8 \times 8$ and $15 \times 15$. 
Among these choices there was little change in the estimates and subsequent analysis uses a $9 \times 9$ window. 
The estimation was performed on the NCAR Cheyenne supercomputer \cite{cheyenne} using the \texttt{R} programming language \cite{Rcore} with the \texttt{Rmpi} \cite{yu2002rmpi} and \texttt{fields} packages \cite{fields}. 
The details of the parallel implementation are the same as in \cite{nychka2018modeling}. Since the fields were standardized, the constraint $\sigma^2 = 1 - \tau^2$ was included. 

The estimates for the spatially varying parameters are shown in Figure \ref{f:moving}. 
The variance and nugget are shown in (a) and (b). 
Panel (c) shows the geometric mean of $\xi_{s_1}$ and $\xi_{s_2}$ as a measure of the average correlation range, and this also agrees with the range in the isotropic case. 
Finally in panel (d), the estimated anisotropy matrix $A(\mathbf s_i)$ is depicted by glyphs indicating the range and departure from isotropy. 
The large signal to noise ratio $\sigma^2/\tau^2$ (not shown) and the evident transition in the covariance structure between land and ocean suggests that the non-stationarity in the second-order structure of the data is being accurately estimated. 
Based on the coastlines in some regions, we hypothesize that the addition of a land/ocean covariate may also be useful.

\subsection{Model checking}

\begin{figure}
    \centering 
   \includegraphics[width=5in]{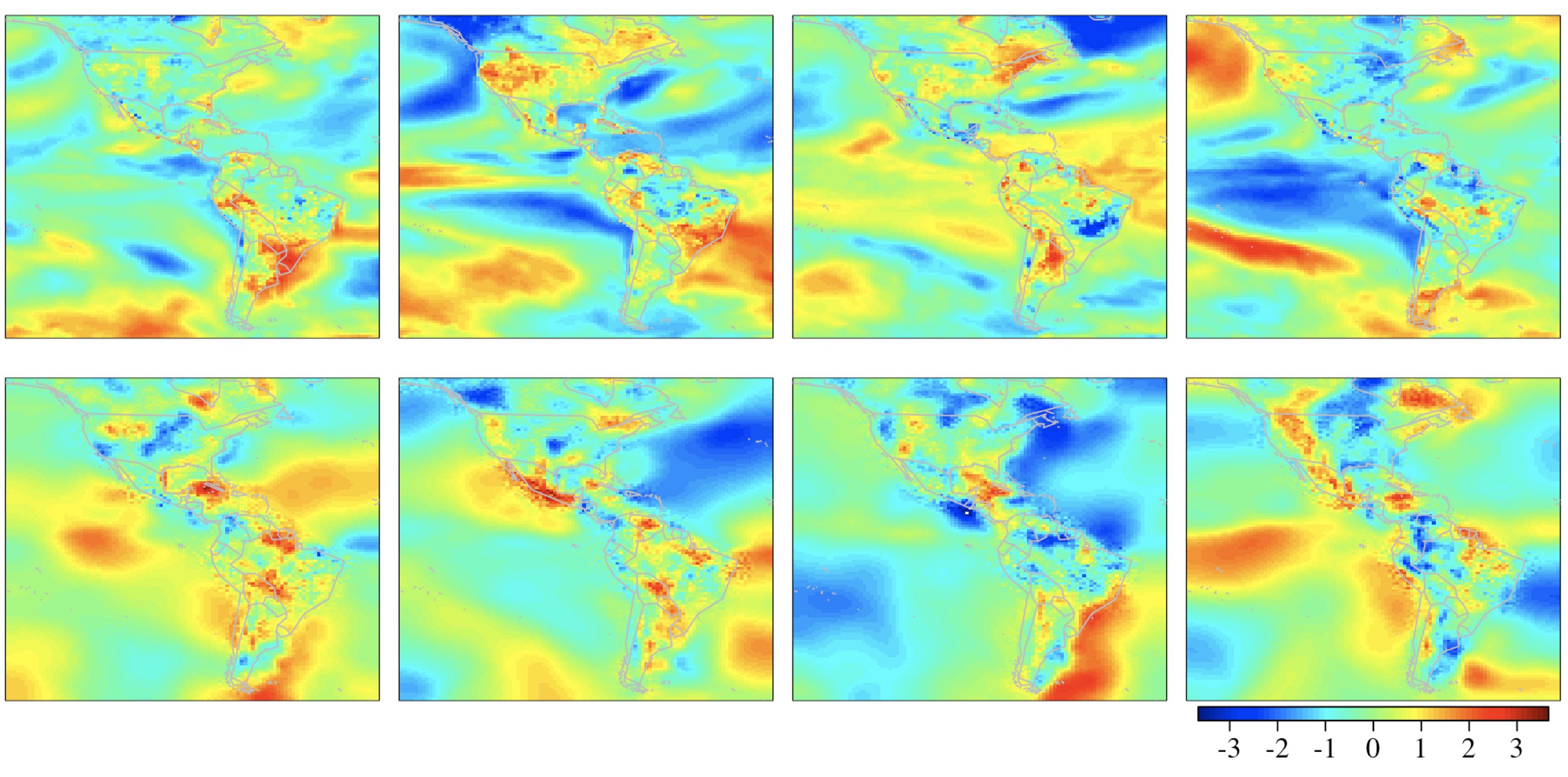}
    \caption{The top row consists of the first four ensemble members from the NCAR CESM data set. The bottom row shows four unconditional simulations from the estimated non-stationary SAR model.}
    \label{f:4}
\end{figure}

\subsubsection{A visual comparison}
The non-stationary SAR model is convenient for simulating high dimensional fields using plug-in estimates of locally varying parameters. 
For this reason, we translate the local Mat\'ern parameters into their approximate SAR parameter equivalents. 
The translation is done using the numerical relationship derived in Section \ref{ss:1}. 
Then, the local SAR parameters are encoded into the non-stationary global SAR model. 
Simulations from this covariance are shown in the bottom row of Figure \ref{f:4}. 
The simulations do a reasonable job emulating the data, but are lacking some of the long range anisotropy over the ocean.

\begin{figure}[t]
    \centering
    \includegraphics[width=5in]{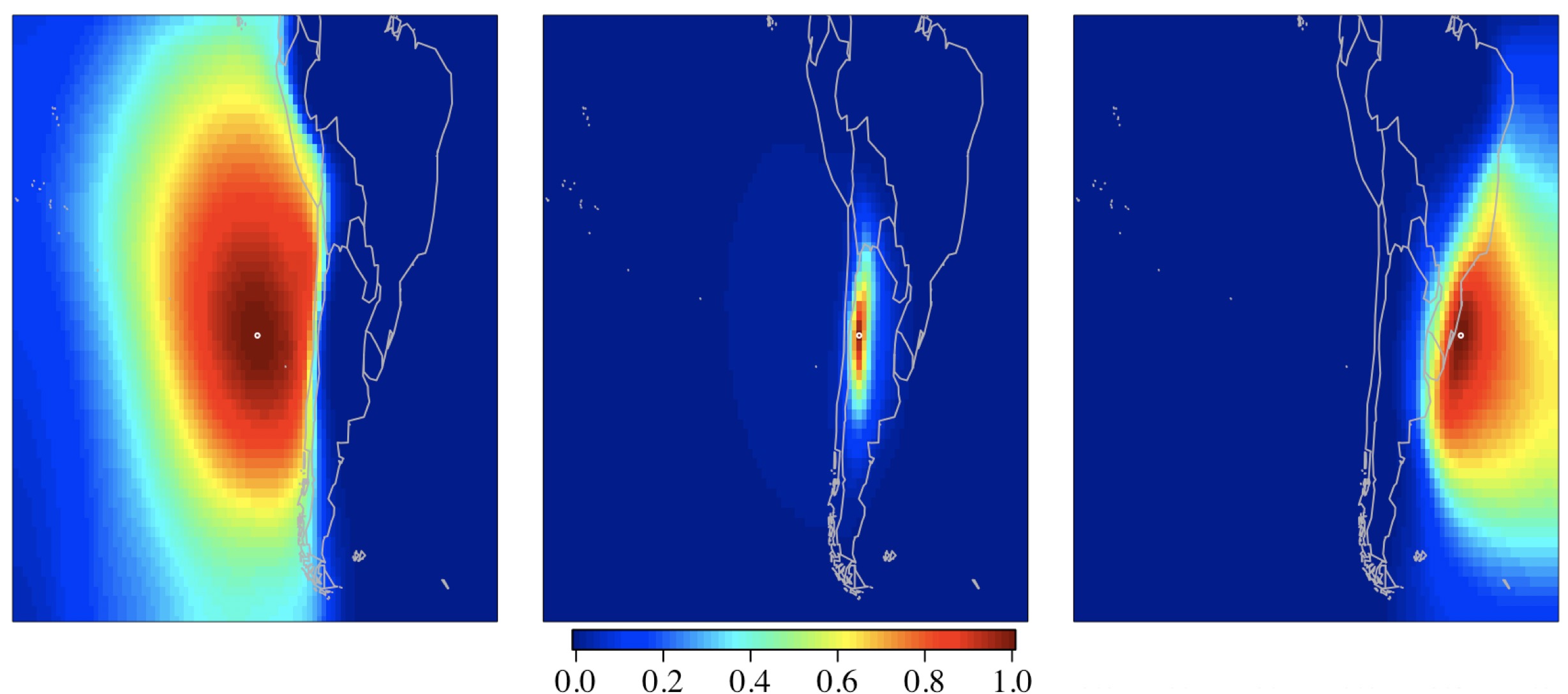} 
    \caption{Estimated correlation functions centered at three locations along the same latitude implied by the non-stationary SAR model.}
    \label{f:5}
\end{figure}

To illustrate how the non-stationarity estimated for this model is related to land/ocean boundaries, Figure \ref{f:5} shows several different locations in the spatial domain and plots the correlations implied by the estimated non-stationary model. 
Both anisotropy and non-stationarity are evident in Figure \ref{f:5}. 

\subsubsection{Transformation to white noise}
The SAR representation as a global model for the spatial field provides a convenient way to check the model fit.   
Under the assumption that the nugget variance is small relative to the smooth Gaussian process, the linear transformation defined by the SAR weight matrix should decorrelate the spatial field. 
In particular, let $\bby$ be the observed field with covariance matrix $\Sigma$. 
The simple idea is to factor $\Sigma$ as $A^{-1} A^{-T}$ and then check that $A\bby$ is a white noise field, or at least a spatial process with greatly reduced spatial dependence. 
Note that the choice of $A$ is not unique, but it makes sense to choose a version of the square root that has weights that are localized around each observation location. 

In this analysis, 
\[
  \Sigma=  \sigma^2 B^{-1} B^{-T} + \tau^2 I  = 
  \sigma B^{-1}( I + \tau^2 Q ) \sigma B^{-T},
\]
where $Q$ is the precision matrix. 
If $\tau= 0$ then the SAR matrix provides a transformation to white noise that is justified, and its approximate will be due just to the local stationary assumption in the fitting and the translation from the Mat\'ern to the SAR, which are both negligible as we have shown.
If $\tau^2$ is small relative to $\sigma^2$, then $B\bby$ will have covariance $(I +  \tau^2Q)$ and will approximate a white noise field. 
Small $\tau^2$ is often  a reasonable assumption in practice because one is often interested in simulation and prediction of spatial data that has strong spatial coherence. 
Note that $Q$ is sparse and when viewed as a covariance matrix will have localized, finitely supported correlations. 

\begin{figure}[t]
    \centering
    \includegraphics[width=5in]{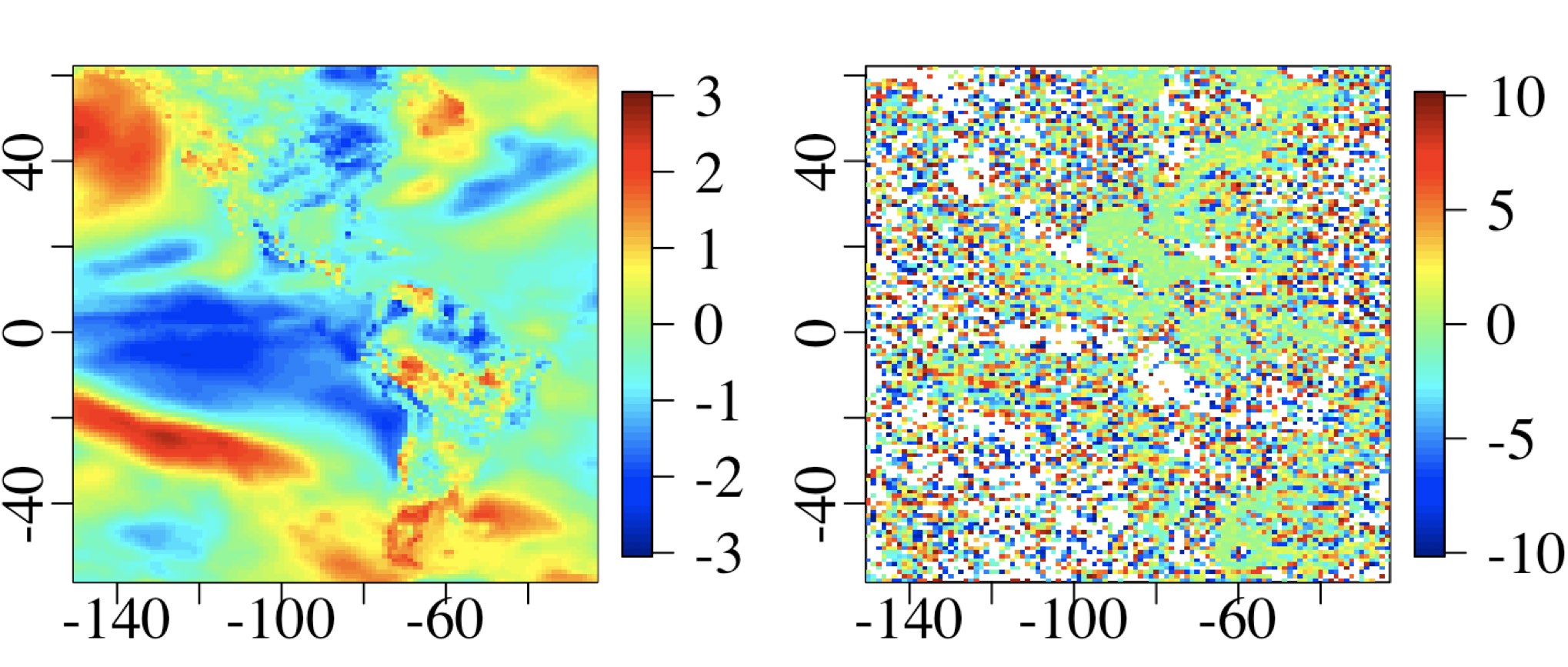} 
    \caption{Panel (a) shows a data replicate $\mathbf y$, and panel (b) shows the symmetric square root of the precision matrix $B$ applied to the data replicate yielding $\mathbf w = B \mathbf y$. }
    \label{f:6}
\end{figure}

Figure \ref{f:6}(b) depicts the result of $B$ applied to one of the replicates (shown in (a)) to which the model was fitted.

As a diagnostic tool one can visually assess the goodness-of-fit of the model covariance matrix to the spatial distribution of the data using these techniques. 
If the spatial distribution of the data is fitted accurately, this process should result in a decorrelated field of white noise. 
Excluding the slight heteroskedasticity present near coastal regions, Figure \ref{f:6} indicates that the vast majority of the correlation in the data has been captured in the model, and therefore has been removed from the data via the matrix transformation. 
This success is encouraging given the long range correlations over the ocean that have been identified from local estimation, and as a SAR only operates on second-order neighbors.  
A formal test of independence was not implemented on the decorrelated fields, although this could be envisioned as a more general goodness-of-fit test in covariance modeling.

\section{Conclusion}

In this paper, we investigate a two-step framework of local estimation and global encoding to represent large and non-stationary spatial datasets. 
We have shown that when independent replicates of spatial processes are available where locally stationarity holds, local maximum likelihood estimation is a robust technique for estimating spatially-varying covariance parameters. 
In particular, the Monte Carlo results indicate the climate model example falls within this context. 

We also explored the stationary Mat\'ern-SAR covariance model approximation, conducting a numerical experiment to compare against existing results. 
The analytic approximation between the models is not reliable for long correlation ranges; however, we can use a numerical approximation to translate parameters between the Mat\'ern and SAR models more accurately. 
To our knowledge this is the first time detailed numerical mappings have been made between the anisotropic SAR model and an anisotropic Mat\'ern covariance function. 

A major contribution of this work is in connecting the local likelihood estimation techniques to a flexible and computationally efficient spatial statistical model based on graphical models. 
We focus on encoding the locally estimated parameters in the non-stationary SAR model. 
In addition, the multistage approach is computationally efficient and can be applied to very large spatial data sets: local estimation avoids the big $n$ problem of global estimation, and encoding local estimates in a SAR model allows us to exploit sparsity for prediction and simulation.
 
Another  advantage of this method is that it can be applied to both continuously indexed and lattice data. 
To reduce the scope of this work we have focused on lattice data. 
Although this restricted format will continue to be standard for climate models, the SAR model can also be extended to irregularly spaced data. 
One approach for non-lattice spatial data is the LatticeKrig model that imposes the SAR and lattice structure on coefficients in a basis function expansion rather than directly on the field.  
We believe the anisotropic models developed here will carry over for more general models such as basis expansions, and the inverse square root transformation will be an important diagnostic tool for non-stationary modeling. 

\section{Acknowledgements}
This research was supported in part by NSF Awards {\tt DMS-1811294} and {\tt DMS-1923062}. Nychka also thanks {\it The International Environmetrics Society (TIES)} for the generous support to attend the 28th Annual TIES Conference, July, 2018
in  Guanajuato, Mexico and to give the President's Invited Lecture. This article constitutes some of  the core research material from that lecture. Finally, the authors acknowledge Stacey Alexeeff and Claudia Tebaldi for formulating the climate scaling application and creating the original data set from LENNS. 

\bibliography{Nonstat.bib}

\end{document}